\documentclass[aps,prl,showpacs,twocolumn,superscriptaddress]{revtex4}
\usepackage{bm}
\usepackage{epsfig}
\usepackage{times}
\usepackage{amssymb}
\usepackage{subfigure}

\begin{document}

\title{Strong Orientation Effects in Ionization of H$_2^+$ by Short,
 Intense, High-Frequency Light Sources}

\author{S. Selst\o}
\affiliation{Department of Physics and Technology, University
  of Bergen, N-5007 Bergen, Norway}
\author{M. F\o rre}
\affiliation{Department of Physics and Technology, University
  of Bergen, N-5007 Bergen, Norway}
\author{J.~P. Hansen}
\affiliation{Department of Physics and Technology, University
  of Bergen, N-5007 Bergen, Norway}
\author{L.~B. Madsen}
\affiliation{Department of Physics and Astronomy, Aarhus University,
DK-8000 Aarhus C, Denmark}

\begin{abstract}
We present three dimensional time-dependent calculations of
ionization of  arbitrarily spatially oriented H$_2^+$ by
attosecond, intense, high-frequency laser fields. The ionization
probability shows a strong dependence on both the internuclear
distance and the relative orientation between the laser field and
the internuclear axis.
\end{abstract}
%

\maketitle
The ionization dynamics of one- and two-electron processes in
diatomic molecules in short, strong laser fields are at present
under intense experimental
investigation~\cite{Ullrich,Cooke,Urbain}. A part of these
investigations  also focus on the sensitivity of such processes to
molecular orientation with respect to the light
polarization~\cite{Litviniuk}. This is again related to the
ultimate goal of controlling chemical reactions by aligning the
reactive molecules with respect to each other prior to the
intermolecular interaction~\cite{Stapelfeldt}.

From a theoretical viewpoint such studies are extremely complex in
the strong-field regime and have been of continuous interest for
nearly two decades (for reviews, see,
e.g.,~\cite{BandraukPosthumus}). In general, only results based on
approximate theories such as the molecular strong-field
approximation~\cite{Becker,Lars1} and tunneling~\cite{Lin} models
have been applied to calculate effects related to molecular
orientation with respect to the light polarization vector. Such
approximate theories are, however, often gauge
dependent~\cite{Lars1,Lars2} and limited in their applicability to
describe complex processes. The "slowness" of past and present
computers have, combined with computational challenges related to
Coulombic multi-center problems, restricted exact theoretical
calculations including both electronic and nuclear degrees of
freedom to cases where the internuclear axis is parallel with the
linear polarization direction~\cite{Dundas, McCann} or models of
reduced dimensionality~\cite{Taeb, Thum, Roudnev}. These studies
have given insight into the fascinating interplay between
electronic and nuclear degrees of freedom; phenomena which at
present are beyond reach of full dimensional computations.

In this Letter, we present the first full time-dependent three
dimensional calculations for the electronic degrees of freedom in
H$_2^+$ exposed to a short, strong, attosecond laser pulse. The
purpose is to follow the behavior of the system with internuclear
distance and in particular to display the dependence of the
dynamics on the angle between the internuclear axis and the linear
polarization of the field. Calculations are performed for 6 cycle
pulses with $\omega=2$ a.u. ($23$ nm) central frequency. This
corresponds to pulse durations around 450 as, which have already
been demonstrated~\cite{atto}. The ionization probability for
H(2p) atoms exposed to similar light sources~\cite{2pArt} showed a
factor 10 stronger modulation with changing orientation than what
was measured with femtosecond pulses~\cite{Litviniuk}. Similar
effects in diatomic molecules may thus indicate that attosecond
pulses may be sensitive probes of the internal nuclear quantum
state as well as its orientation. The calculations  indeed display
that the ionization probability depends strongly on these
parameters. Atomics units ($\hbar=m_e=e=1$) are applied
throughout.

As the nuclear vibrational period is approximately $10^3$ times
larger than the pulse duration, the nuclear degrees of freedom can
be considered frozen during the attosecond pulse. Post pulse
interplay between nuclear and electronic degrees of freedom, which
are important for weaker fields, are also found to be unimportant
here as direct electronic ionization dominates.

The vector potential for the light source  is given by
\begin{equation}
\label{Afelt} {\bf A}(t)=\frac{E_0}{\omega} \sin^2\left(
\frac{\pi}{T} \, t \right) \, \sin (\omega t + \phi) \, {\bf u}_p,
\end{equation}
where ${\bf u}_p$ is a unit vector defining the orientation of the
linearly polarized field, and  $\phi$ is chosen such that the
field corresponding to  (\ref{Afelt}) represents a physical
field~\cite{Lars2}. The validity of the dipole approximation was
investigated in detail very recently for the present intensity and
frequency regime, and was found to be well-justified for
ionization~\cite{newhamiltonian}. The vector potential determines
the electric field, ${\bf{E}}(t)= -\partial_t {\bf A}(t)$, and the
translation, ${\bm \alpha} (t) = \int_0^t{\bf A}(t') \, dt'$, which
enter the length $H_l$ and the Kramers-Henneberger $H_{K\!H}$
form~\cite{PKH} of the interaction Hamiltonian, respectively,
\begin{equation}
\label{Hl} H_l=\frac{p^2}{2}-\frac{1}{|{\bf r}+ {\bf R}/2|}-
\frac{1}{|{\bf r}- {\bf R}/2|}+ {\bf E}(t) \cdot {\bf r},
\end{equation}
\begin{equation}
\label{Hv} H_{K \! H}= \frac{p^2}{2}-\frac{1}{|{\bf r}+ {\bf R}/2
+\mbox{\boldmath $\alpha$}(t)|}- \frac{1}{|{\bf r}- {\bf
R}/2+\mbox{\boldmath $\alpha$}(t)|}
\end{equation}
with ${\bf R}$ the internuclear distance. Both versions of the
Hamiltonian have been applied here to secure invariant results.

For fixed nuclei, we solve the time-dependent Schr{\"o}dinger
equation numerically based on a split-step operator approximation
on a spherical grid. The method was described in detail
elsewhere~\cite{HermannFleck, JanPetter}. The wave function is
expanded on the grid points $\left[(r_i, \Omega_{jk})
=(r_i,\theta_j, \phi_k)\right] $ ,
\begin{equation}
\Psi(r_i, \Omega_{jk},t; {\bf R}) = \sum_{l,m}^{l_{max},m_{max}}
f_{l,m}(r_i,t; {\bf R})Y_{l,m}(\Omega_{jk})
\end{equation}
with origin at the center of the internuclear axis, and with
parametrical dependence on ${\bf R}$. The field-free initial state
$| \Psi_0 \rangle$ is obtained by the substitution $t \rightarrow
-i \, t$ in the propagator. At internuclear separation $R=2$ a.u.
this gives an electronic ground state energy,
$\varepsilon_0^{grid} =-1.099 $ a.u., which can be compared to the
exact value of $\varepsilon_0 = -1.103$ a.u. Reflections at the
edges $r=r_{max}=60$ a.u. are avoided by imposing an absorbing
boundary. We include up to $l_{max} = 15$, $N_r=1024$
radial points, and in the propagation we use $\Delta t = 5 \times
10^{-3}$ a.u. Here we stress that for a given $l_{max}$,
specified at input, the quadrature rule of the spherical harmonics
uniquely fixes the angular points.

At the end of the pulse, $t=T$, a fraction of the wave function
has been removed by the absorber. Since excitation is found to be
a minor channel, the ionization probability  can be calculated as
$P_{ion} =  1- |\langle \Psi_0| \Psi(T) \rangle|^2$.

\begin{figure}
\begin{center}
\epsfysize=5.0cm \epsfbox{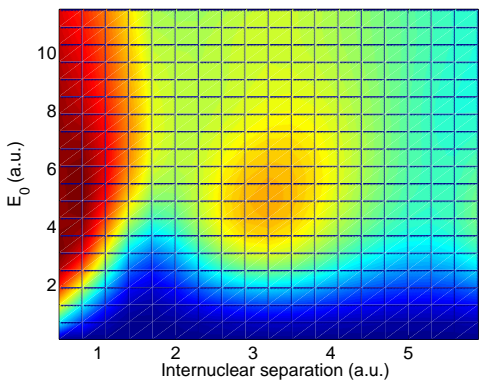}
\end{center}
\caption{\label{fig1} Ionization probability in the parallel
geometry ($\theta=0^\circ$) as a function of the internuclear
separation $R$ and the electric field strength $E_0$ with $\omega=2$
a.u. and $T=6 \pi$ a.u.}
\end{figure}

The ionization probability as a function of internuclear separation
and electric field strength is displayed in Fig.~\ref{fig1} for
field polarization parallel with the internuclear axis. Two
striking maxima are observed, one for small internuclear
separation, $R \sim 1$ a.u.,  and another for $R \sim 3$ a.u. When
the field strength is further increased, the ionization
probability decreases, i.e., the molecule is partly stabilized in
the intense field. This rather counterintuitive mechanism has been
studied in detail for atoms~\cite{Gavrila}. What happens, is that
for increasing intensity, the  final channels corresponding to
single and multiphoton ionization close while shake-off dynamics
with characteristic low-energy electron emission becomes the
dominating ionization mechanism~\cite{newhamiltonian}. Between
these two regions a "valley" in the ionization curve with
increasing intensity may occur as indicated around $R\sim 1$ a.u.
and $R\sim 3$ a.u. for field strengths beyond $E_0 \approx 7$ a.u.
At the ground state equilibrium distance, $R \sim 2$ a.u., and for
$R \sim 5$ a.u., the ionization probability is significantly
smaller, indicating strong dynamic self-interference effects of
the electronic charge clouds associated with each scattering
center.

\begin{figure}
\begin{center}
\epsfysize=5.0cm \epsfbox{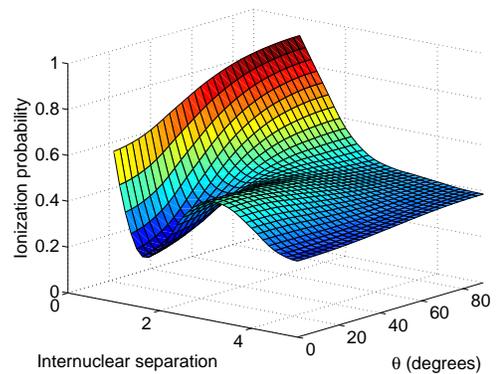}
\end{center}
\caption{\label{fig2} Ionization probability as a function of the
angle $\theta$ between the polarization direction and the
internuclear axis with $\omega=2$ a.u., $E_0=3$ a.u. and $T=6 \pi$
a.u.}
\end{figure}
From Fig.~\ref{fig1} we see that the variation in the ionization
signal is most pronounced for $E_0 \sim 3$ a.u. At this field
strength, Fig.~\ref{fig2} exposes the ionization probability as a
function of internuclear separation and as a function of the angle
$\theta$ between the internuclear axis and the polarization
direction of the field. An oscillatory behavior of the ionization
probability in the parallel geometry ($\theta = 0^\circ$) is seen.
As $\theta$ increases the oscillations gradually decrease, and in
the perpendicular geometry ($\theta=90^\circ$), the ionization
probability drops monotonically with $R$. In the figure, we also
observe opposite functional dependence with $\theta$: At $R \sim 2$
a.u. the ionization probability increases with $\theta$ while at $ R
\sim 3$ a.u. it decreases.

\begin{figure}
\begin{center}
\begin{tabular}{l r}
\multicolumn{2}{l}{ \epsfxsize=5cm \epsfysize=2.5cm \leavevmode
\epsfbox{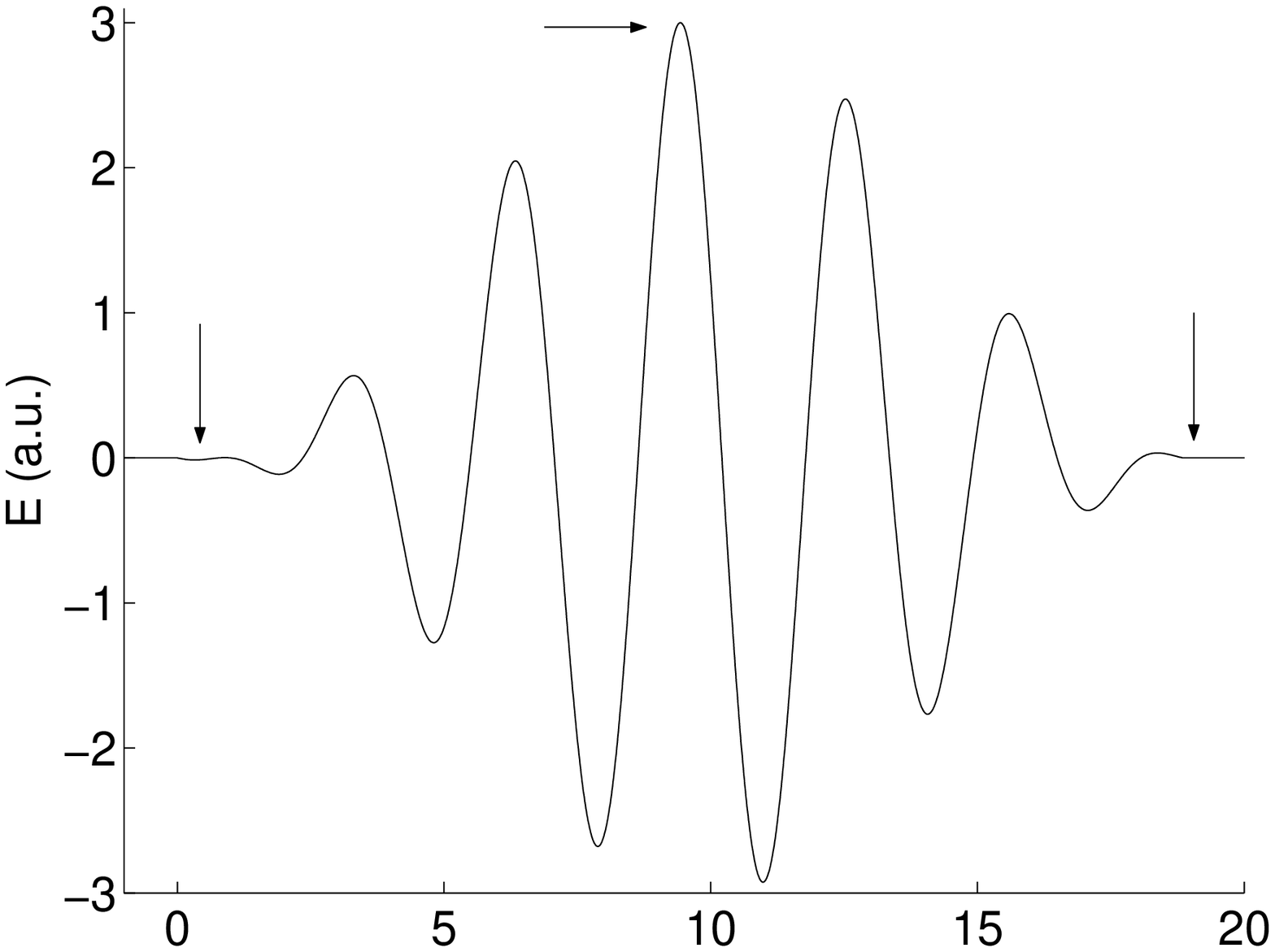}
} \\
\epsfxsize=2.5cm \leavevmode \epsfbox{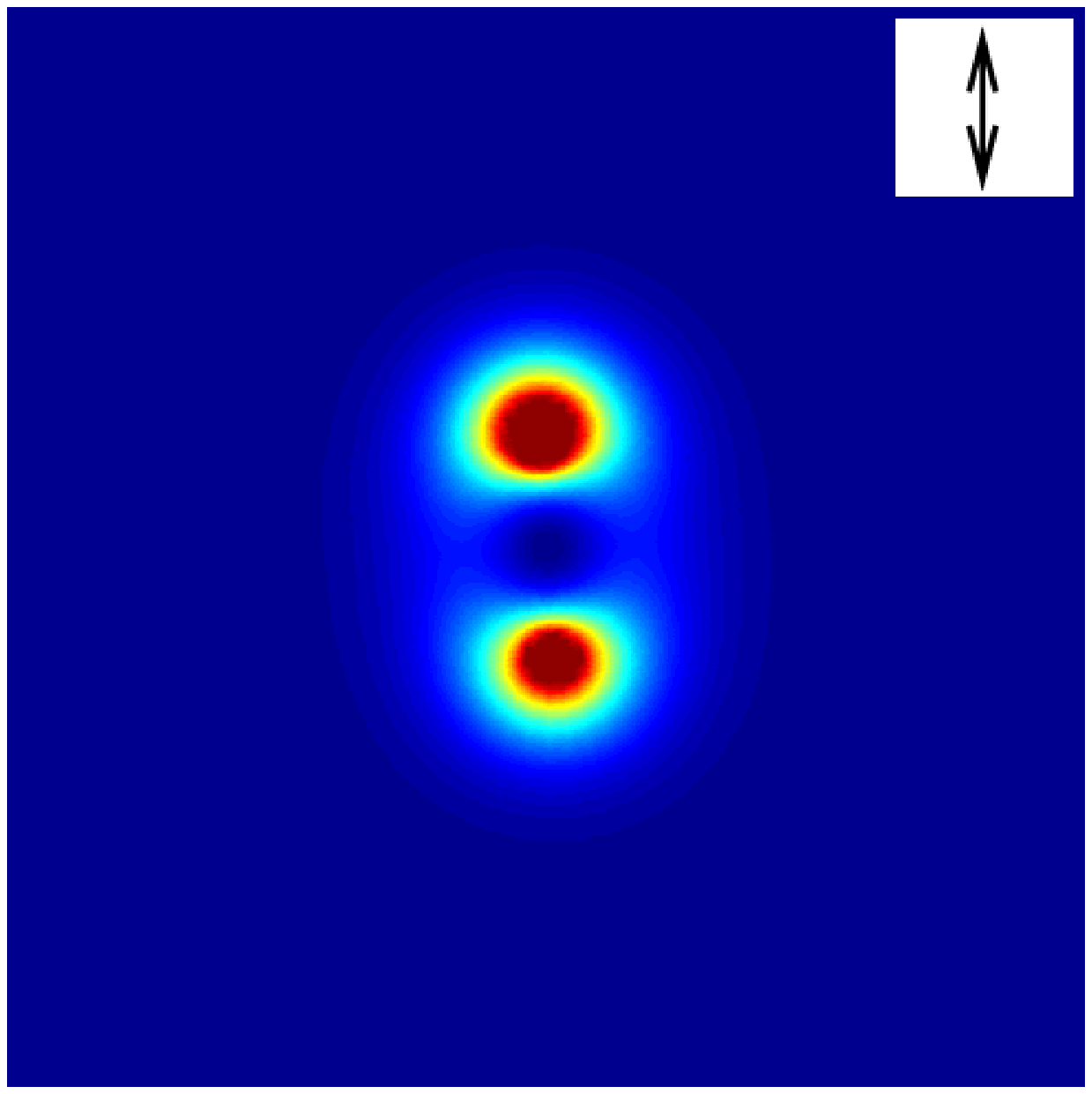} &
\epsfxsize=2.5cm \leavevmode
\epsfbox{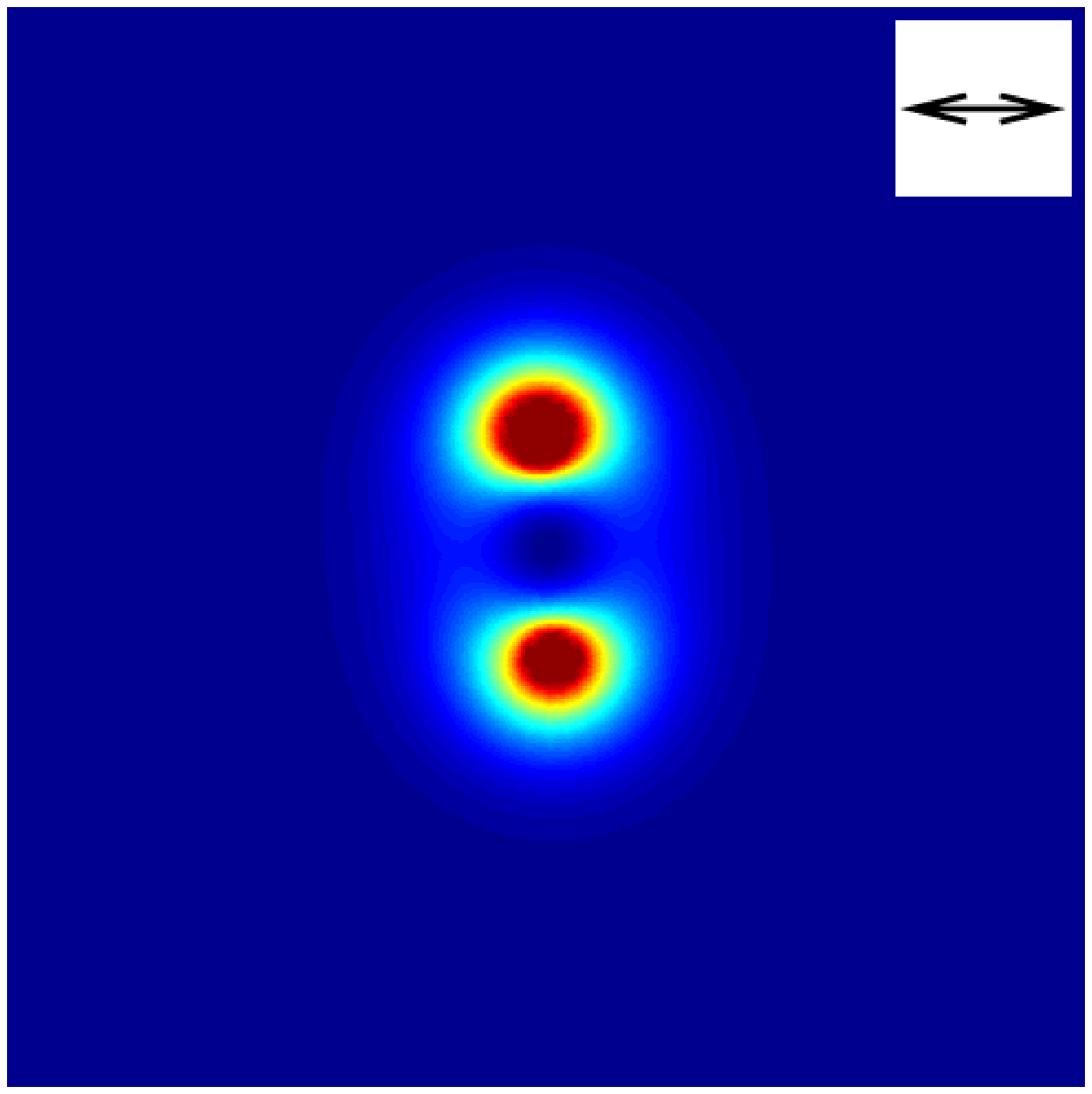} \\
\epsfxsize=2.5cm \leavevmode \epsfbox{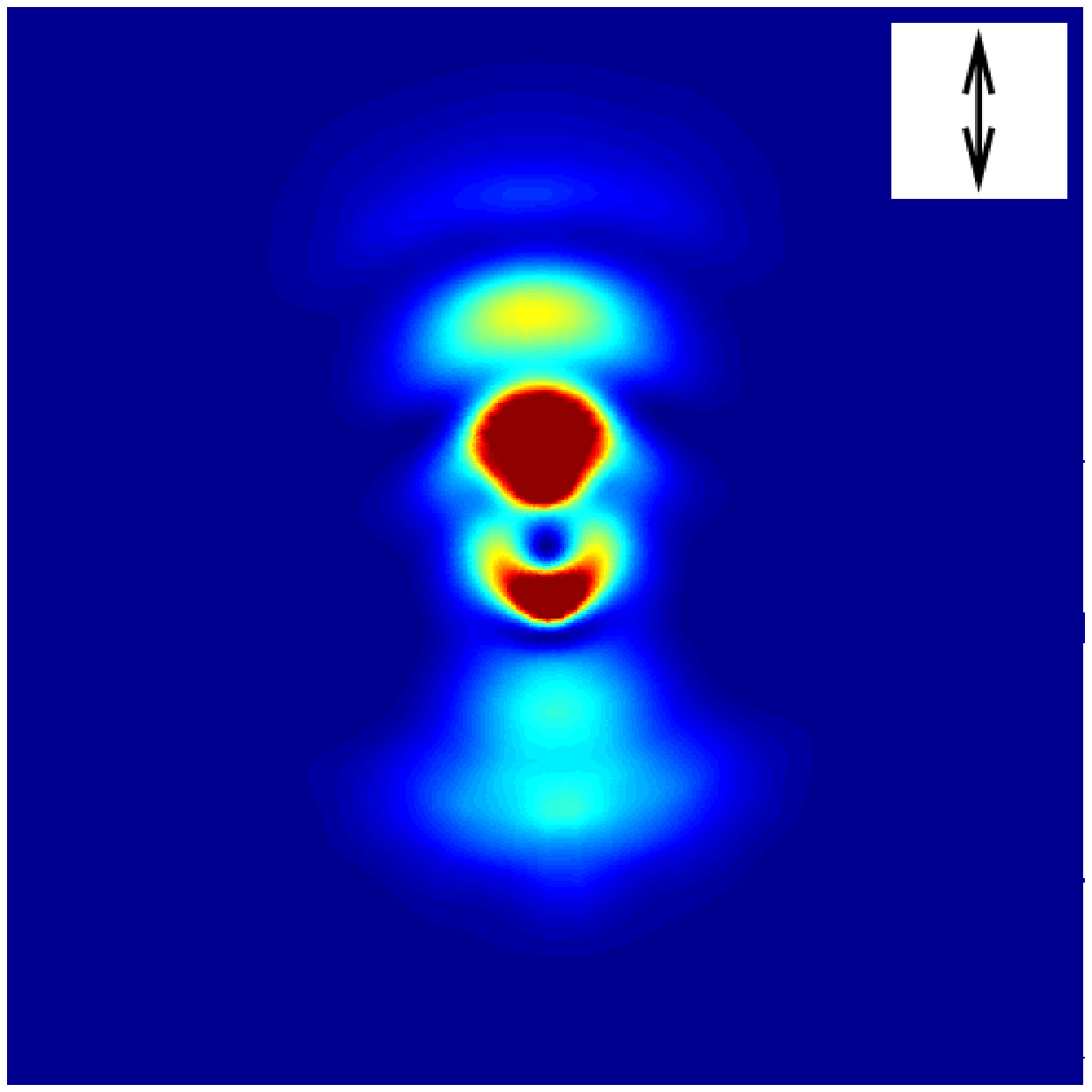} &
\epsfxsize=2.5cm \leavevmode
\epsfbox{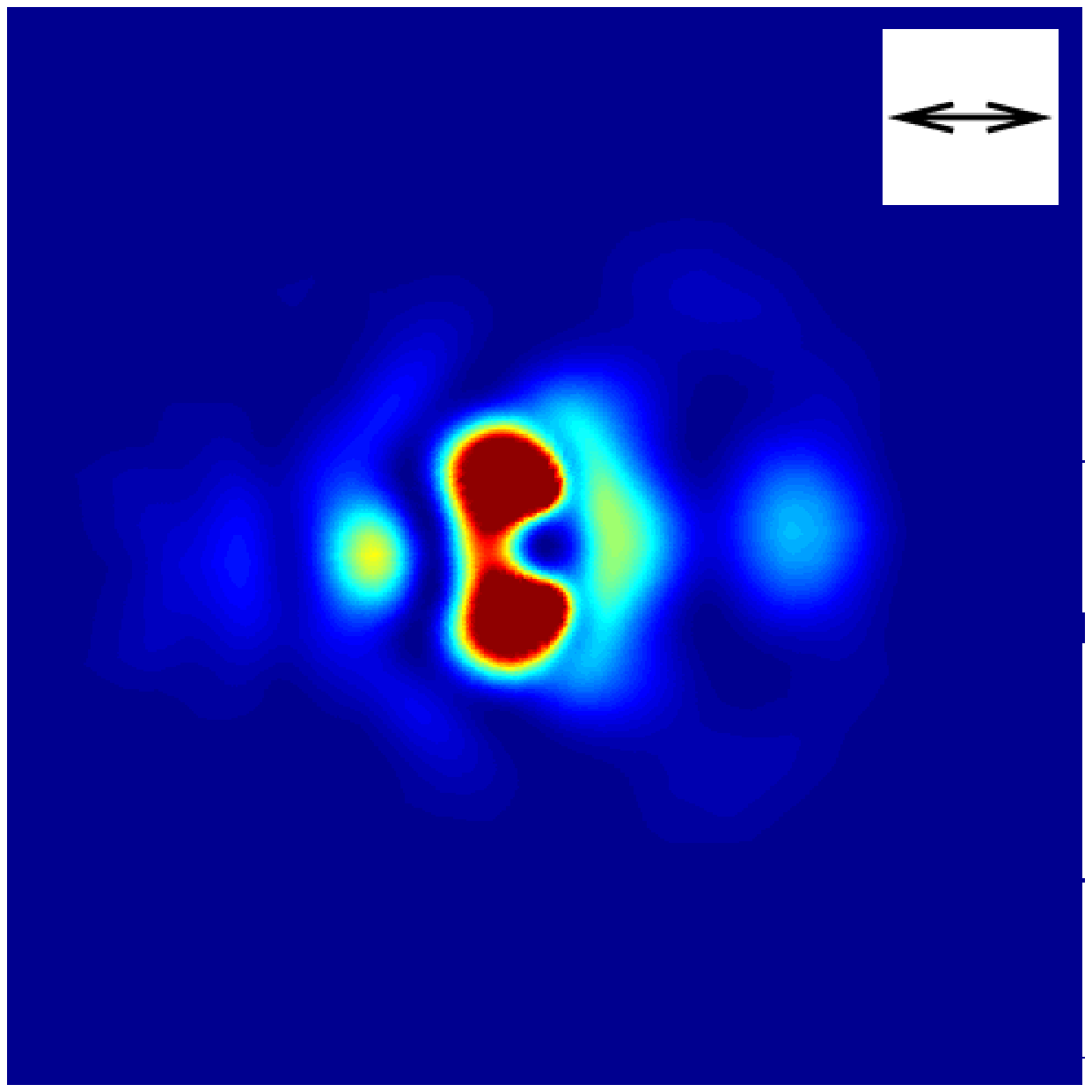} \\
\epsfxsize=2.5cm \leavevmode \epsfbox{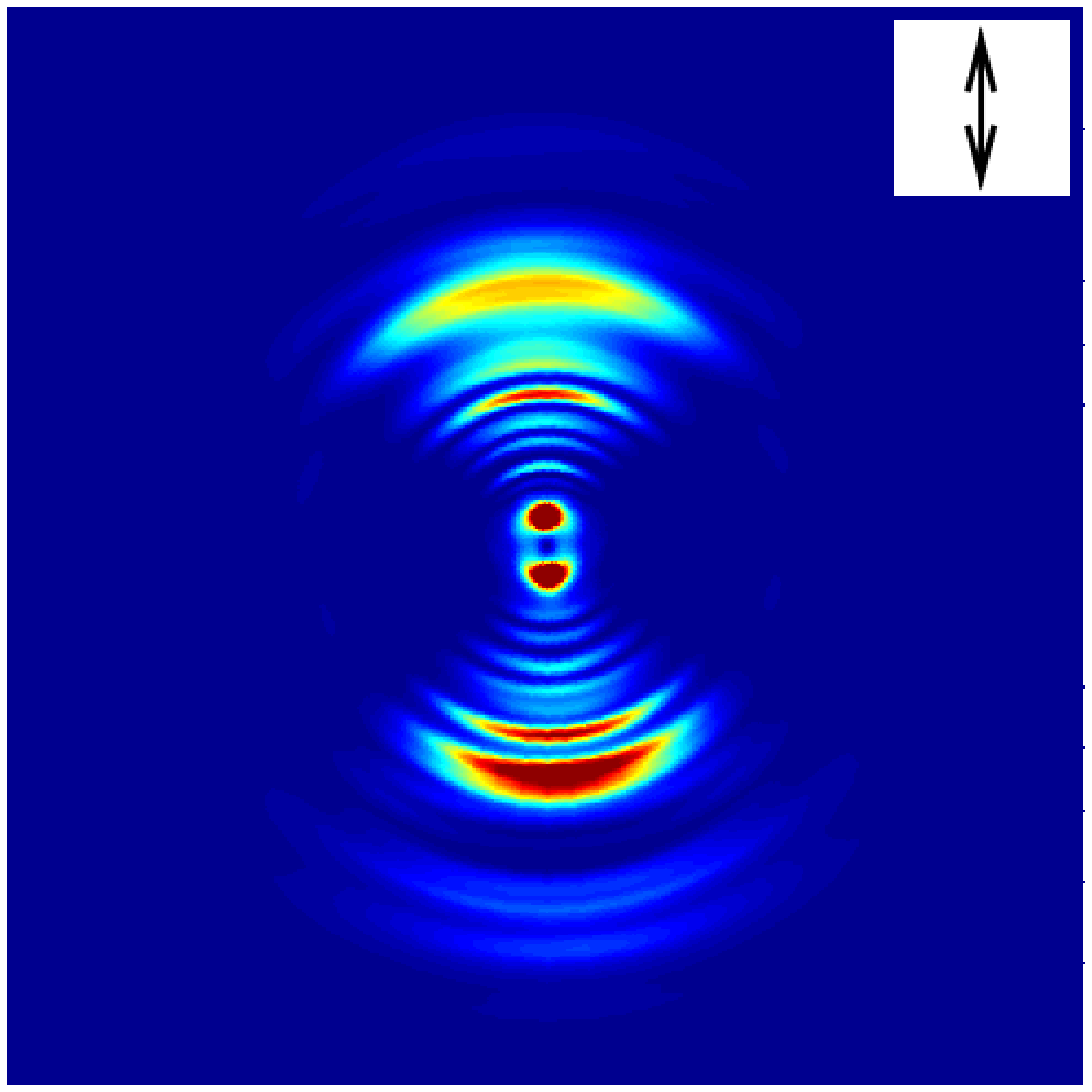} &
\epsfxsize=2.5cm \leavevmode \epsfbox{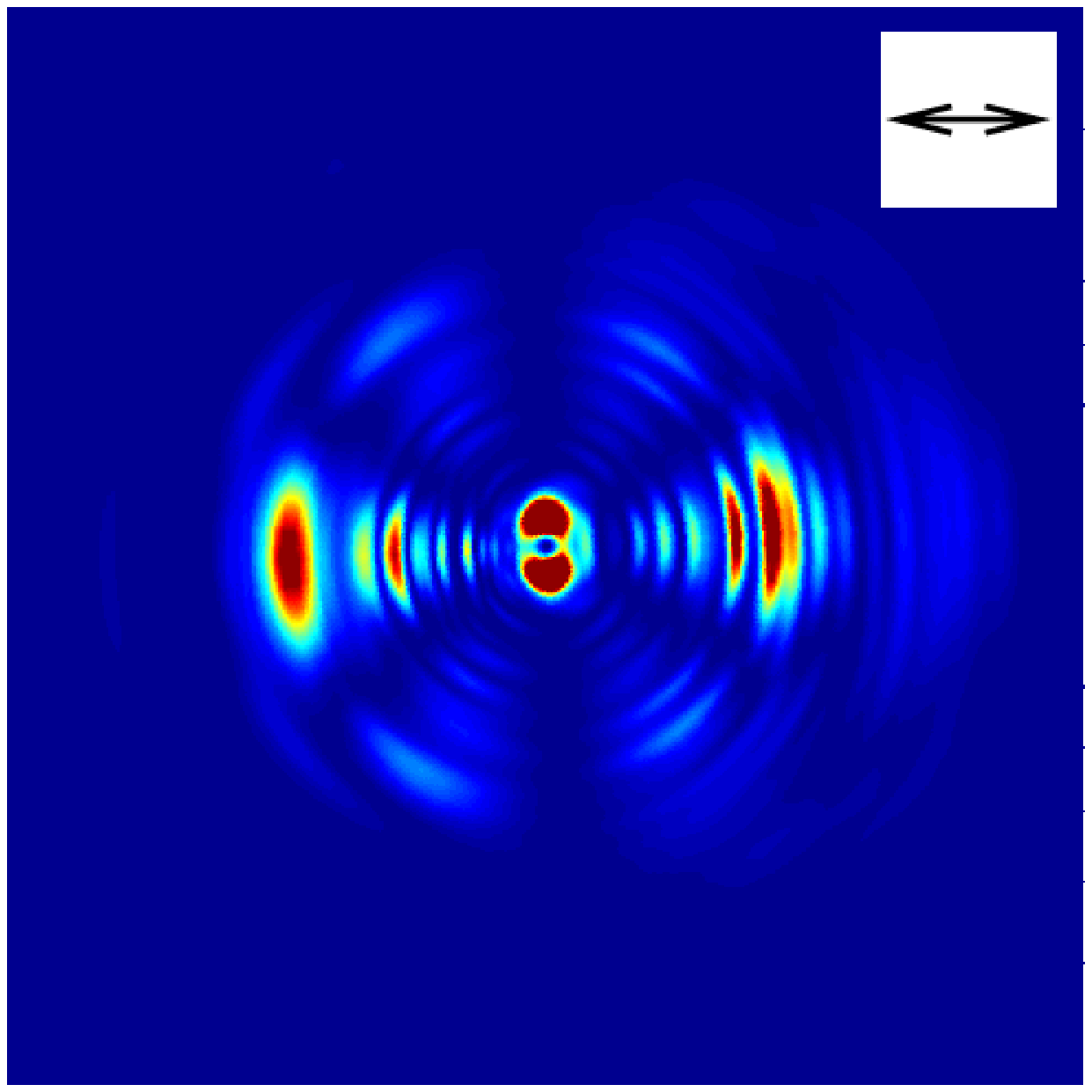}
\end{tabular}
\end{center}
\caption{{\it Top:} The electric field $E(t)$ of duration $T= 6 \pi$
a.u. (450 as) and frequency $\omega=2$ a.u. The arrows indicate the
instants of time at which the snapshots of the lower part of the
figure are made. {\it Color online}: Snapshots of the wave function
in the $xz$-plane at times corresponding to the beginning (top), the
middle (middle) and the end (bottom) of the pulse for parallel
(left) and perpendicular (right) orientation. In all cases the
internuclear separation is $R=3$ a.u. Both the polarization
direction and the internuclear axis lie in the $xz$-plane.}
\label{fig3}
\end{figure}

We now turn to the detailed dynamics and a qualitative
understanding of the processes underlying the phenomena observed
in Fig.~\ref{fig2}. Figure~\ref{fig3} shows snapshots of the wave
function in the $xz$-plane at various times for parallel and
perpendicular polarization (the molecule has its internuclear axis
directed along $z$). In general the photoelectron is ionized in
the directions of the field. For $\theta = 0^\circ$ the initial
charge cloud is partly dragged back and forth along the field, and
this gives rise to a strong interference between various momentum
components of the wave function and hence the oscillatory
dependence with $R$ in Fig.~\ref{fig1}. This effect is absent at
$\theta = 90^\circ$ where the two atomic-like charge clouds
pertaining to each nucleus oscillates in phase back and forth with
the electric field. In the lower right panel, secondary intensity
maxima appear at $30^\circ$ and $150^\circ$ with respect to the
internuclear axis. These structures look similar to double slit
scattering but have a more subtle dynamical origin.

The oscillatory behavior in $P_{ion} (R)$ was discussed in a one
dimensional model earlier in terms of the concept of enhanced
ionization~\cite{Seideman}. Related phenomena were also discussed in
fast ion-molecule collisions~\cite{collisions}. The following simple
{\it ansatz} gives a qualitative explanation of the oscillations and
their absence at $\theta=90^{\circ}$: Assume that the outgoing waves
from each of the scattering centers are a superposition of two
outgoing spherical waves,
\begin{equation}
\psi_{out}= f_1(\Omega_1) \frac{e^{ik|{\bf r}+ {\bf R}/2|}} {|{\bf
r}+ {\bf R}/2|} + f_2(\Omega_2) \frac{e^{ik|{\bf r}- {\bf R}/2|}}
{|{\bf r}- {\bf R}/2|}.
\end{equation}
If we take the two scattering amplitudes to be equal,
$f_1(\Omega_1)=f_2(\Omega_2)$, the differential ionization
probability can be brought to the form
\begin{equation}
d P_{ion}/ d\Omega \propto 2 |f_1(\Omega)|^2 \, (1+ \cos(k {\bf
\hat{r}} \cdot {\bf R}))
\end{equation}
for $r \gg R$. As  seen from Fig.~\ref{fig3}, the main part of the
outgoing wave follows the orientation of the field. Hence we
expect that for parallel polarization, the main contributions will
be for ${\bf \hat{r}}$ parallel to ${\bf R}$, which again gives
raise to oscillations in $R$ with wave number $k$. Given that the
one-photon ionization dominates, we find the wave number as $k
\approx \sqrt{2(\omega - \tilde{I_p}(R))}$, where $\tilde{I_p}$ is
the effective ionization potential. This is seen to be consistent
with the results in Figs.~\ref{fig1} and \ref{fig2} and we have
also confirmed these findings for other values of $\omega$. The
absence of oscillations in the case for perpendicular polarization
is understood accordingly: The wave is sent out mainly in the
direction given by $\theta= 90^{\, \mathrm o}$, and since the
outgoing waves will have no phase difference due to the separation
of the scattering centers in this direction, no interference
pattern will occur (${\bf \hat{r}} \cdot {\bf R} = 0$).

The monotonous decrease in $P_{ion}$ with $R$ at $\theta=
90^\circ$ is reasonable since the decrease in the ionization
potential leads to an effective higher final state electronic
momentum with increasing $R$.

\begin{figure}
\begin{center}
\epsfysize=5.0cm \epsfbox{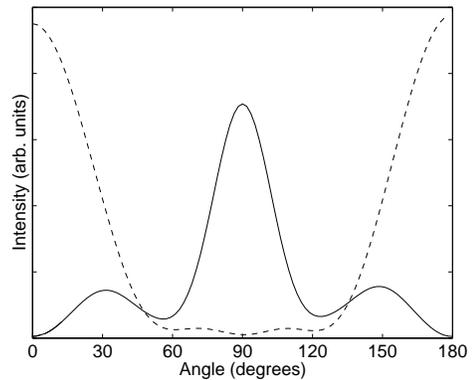}
\end{center}
\caption{Angular photoelectron spectrum in the scattering plane for
parallel (dashed curve) and perpendicular (full curve) geometry as a
function of the polar angle for a 6 cycle field with $E_0=3$ a.u.
and $\omega=2$ a.u. In both cases, the angle denotes the direction
of the outgoing electron with respect to the internuclear axis. We
have used equal normalizations for the two curves.}\label{fig4}
\end{figure}
The angular distribution of the ionization probability can be
calculated from the time integral of the radial current density
through the solid angle element $d \Omega$ at a chosen distance $a$
from the origin
\begin{equation} \label{dPdTh} \frac{d P_{ion}}{d \Omega}= \int_0^\infty
\! dt \; {\bf j}(a,t) \cdot {\bf \hat{r}} =  \int_0^\infty \! dt
\;\Im {\mathrm m} \left( \left. \Psi^* \frac{\partial \Psi}{\partial
r} \right|_a \right),
\end{equation}
where the distance $a$ is chosen large enough to exclude
contribution to the current from the quiver motion of an electron
close to the nucleus and small enough to avoid effects induced by
the absorber. The application of this procedure to the outgoing
waves of the lower panel of Fig.~\ref{fig3} results in the
intensity spectrum of Fig.~\ref{fig4}. As already observed from
Fig.~\ref{fig3}, the photoelectrons are most likely
ejected in the direction of laser polarization. In the
perpendicular arrangement the two peaks around $30^{\circ}$ and
$150^{\circ}$ can  also be quantified: The integrated probability
connected to these secondary maxima amounts to about 25\% of the
total ionization probability. As these peaks originate from
interference of the outgoing waves, it is interesting to note that
in a very recent calculation of High Harmonic Generation (HHG) in
a reduced model with respect to the electronic degrees of freedom
such interference does not occur \cite{Lein}. In that work it is
pointed out that orientational effects are very important for HHG.

In conclusion, fully non-perturbational calculations of the
ionization dynamics of H$_2^+$ molecules in intense attosecond
light sources have been carried out. Very strong orientation
effects have been found,  demonstrating that in order to obtain a
full understanding of the molecular ionization dynamics, all three
electronic degrees of freedom must be included. The qualitative
features are determined by interference effects related to
double-center scattering and the distinct features in the electron
spectra show that intense attosecond pulses can resolve the
instantaneous vibrational and orientational quantum state of
diatomic molecules.

\begin{acknowledgments}
It is a pleasure to thank Thomas K. Kjeldsen for useful
discussions and for critically reading the manuscript. The present
research was supported by the Norwegian Research Council through
the NANOMAT program and the Nordic Research Board NordForsk and
by the Danish Natural Science Research Council.
\end{acknowledgments}

\end{document}